# Title Page

**Classification:**
PHYSICAL SCIENCES / Applied Physical Sciences

**Title:**
Shear Forces and Heat Conductance in Nanoscale Junctions.

**Authors:**
Benjamin. J. Robinson[1,2], Manuel. E. Pumarol[1], and Oleg. V. Kolosov*[1,2]

**Author Affiliations:**
[1] Physics Department, Lancaster University, Lancaster, LA1 4YB, UK
[2] Materials Science Institute, Lancaster University, Lancaster, LA1 4YW, UK

**Corresponding Author:**
Professor Oleg. V. Kolosov
Physics Department
Lancaster University
Lancaster, LA1 4YB
UK
Tel: +44 (0)1524 593619
e-mail: o.kolosov@lancaster.ac.uk

# Shear Forces and Heat Conductance in Nanoscale Junctions

Benjamin. J. Robinson[1,2], Manuel. E. Pumarol[1], and Oleg. V. Kolosov*[1,2]

[1] Physics Department, Lancaster University, Lancaster, LA1 4YB, UK; email: o.kolosov@lancaster.ac.uk

[2] Materials Science Institute, Lancaster University, Lancaster, LA1 4YW, UK

**ABSTRACT:** Nanoscale solid-solid contacts define a wealth of material behaviours from the electrical and thermal conductivity in modern electronic devices to friction and losses in micro- and nanoelectromechanical systems. For modern ultra-high integration processor chips, power electronic devices and thermoelectrics one of the most essential, but thus far most challenging, aspects is the assessment of the heat transport at the nanoscale sized interfaces between their components. While this can be effectively addressed by a scanning thermal microscopy (SThM), which demonstrates the highest spatial resolution to thermal transport to date, SThM's quantitative capability is undermined by the poorly defined nature of the nanoscale contact between the probe tip and the sample. Here we show that simultaneous measurements of the shear force and the heat flow in the probe-sample junction shows distinct correlation between thermal conductance and maximal shear force in the junction for multiple probe-material combinations. Quantitative analysis of this correlation confirmed the intrinsic ballistic nature of the heat transport in the tip-surface nanoscale contact suggesting that they are, ultimately, composed of near-atomic sized regions. Furthermore, in analogy to the Wiedemann-Franz law, which links electrical and thermal conductivity in metals, we suggest and experimentally confirm a general relation that links shear strength and thermal conductance in nanoscale contacts *via* the fundamental material properties of heat capacity and heat carrier group velocity, thus opening new avenues for quantitative exploration of thermal transport on the nanoscale.

## Significance

Understanding the nature of heat transport at the nanoscale, where our classical model of what is heat breaks down, is of supreme importance for a vast range of modern nanoscale devices. We address this by simultaneously studying shear forces, effectively - friction, and thermal conductance in nanoscale junctions that join contacting bodies. The correlation between these two measurements, each having a very different physical origin, allows determination of the nanoscale contact area and experimentally confirms that on the close-to-atomic length scale the heat transport in the contacts is of ballistic nature. Furthermore, this study uncovers a generic relationship linking shear strength and thermal conductance in nanoscale contacts much like the Wiedemann-Franz law which links electrical and thermal conductivity in metals.

## Introduction

As continuously decreasing length scales are being exploited in electronic devices, nanoelectromechanical systems (1) and nanomaterials (2), understanding the true nature of nanoscale contacts between solid surfaces and interfaces is essential. Whilst electron transport (3) and force interaction (4) in such contacts have been well explored, studying nanoscale heat transport in active (heat generating) and passive (heat dissipating) nanostructures (5) still poses significant challenge (6). This is further complicated as the critical dimensions of these modern devices continue to decrease below the mean free path (MFP) of electrons and phonons – the two major types of heat carriers in solid state devices (7).

The leading instrument for nanoscale thermal mapping, scanning thermal microscopy (SThM) (8), relies on the heat transfer between a nanometre dimension apex of a heated thermal probe (9, 10) and the studied sample, the resulting change of the probe temperature on contact with the sample allowing estimation of its local thermal conductivity (11, 12). Unfortunately, due to the generally irregular and fluctuating morphology of the nanoscale solid-solid contact (shown schematically in Fig. 1), these phenomena are difficult to model and even more difficult to experimentally investigate (13) therefore significantly reducing the reliability and effectiveness of nanoscale heat transport measurements in SThM (14).

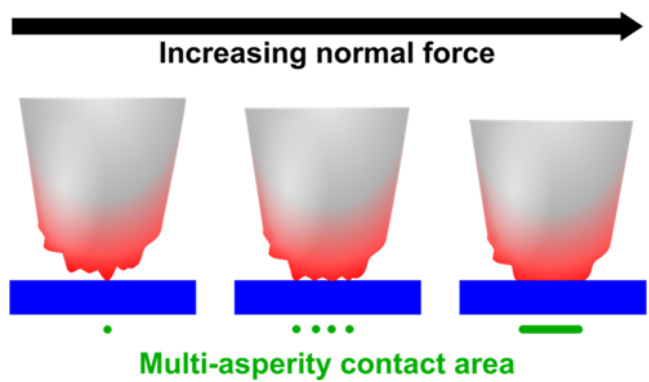

**Fig. 1. Schematic of multi-asperity contact between a heated SThM probe and sample where contact area between probe and sample does not increase linearly (green lines) due to the unpredictable and fluctuating nature of the probe-sample nanoscale interface.**

A tempting approach to resolve this would be to find a parameter complimentary to the heat transfer which is also sensitive to the state of the nanoscale junction. Recently, studies of the heat and electron transport in metallic junctions (15) confirmed that the centuries old Wiedemann-Franz law (16) describing the proportionality of thermal and electrical conductance in metals is correct down to the atomic scale. Another study comparing dependence of the thermal transport in the nanoscale junction with the normal forces (13) allowed the exploration of the effect of nanoscale and atomic roughness on the heat flow in these junctions; however in this study, the contact area, the physical value that ultimately defines the contact, was only indirectly determined *via* the generally unknown tip shape. To determine this area one could measure a tip-surface compliance, a value that directly depends on the contact area (17), but this would require reliable determination of high contact stiffness values on the order of several 1000 N m$^{-1}$. Some probe approaches based on high frequency ultrasonic vibrations (18), are sensitive to this range of stiffness values, but offer only a semi-quantitative solutions and therefore cannot be used in this study.

Here we used another physical parameter that is both easily measurable in the scanning probe microscope and also known to be directly linked with the contact area of a junction – namely, the maximal shear force the junction can support, here called the "shear response". It was shown, by detailed nano-tribology studies elsewhere (19-22), to be directly proportional to the solid-solid contact area of the junction. In our experiments, this shear response is acquired *via* lateral dithering of the sample perpendicular to the long axis of the cantilever while measuring the maximum shear

(friction) force $F_{ms}$ during such dithering (*Methods*). The high torsional stiffness of the SThM probe suggests that the variation of the tip-surface distance across the junction is several orders smaller than an inter-atomic distance and hence should produce no effect on the heat flow in the junction; this was also experimentally confirmed during the study (see Methods). We then used a change in a normal force, $F_n$, e.g. during approach and retract of the tip to the probed surface, at rate much slower than dithering frequency, to modulate the contact area while measuring concurrently both a change in the maximal shear force $F_{ms}$ as well as the heat conductance of the junction $\sigma$ in the SThM setup.

Unlike comparisons of electrical and thermal transport (15, 23), both shear force and heat flow are present for any metallic, semiconductor or insulating contacts, hence this study provides a universal platform for exploring fundamental heat transport phenomena in solid-solid nanoscale junctions (24).

**Experimental Results**

**Correlation of nanoscale thermal and shear responses:** In our experiments, as described in *Methods*, we used a Joule self-heated micro-fabricated resistive probe. The change of probe resistance, included in a sensitive Wheatstone bridge circuit, reflected a change of the probe temperature. For a constant power applied to a probe, this temperature (and hence a value of the electrical signal from the bridge – the "thermal response") is a linear function of the probe thermal resistance $R_j$. In absence of the tip-surface contact, the probe thermal resistance is defined by the heat flowing to the base of the probe, but as the nanoscale apex of the probe approached and contacted the sample, an additional heat flow is carried to the sample cooling the tip and changing the probe thermal resistance (25).

The probe thermal response was measured simultaneously to the normal and shear forces during tip-surface approach with typical dependencies shown in Fig. 2. The lateral force (shear) response of the probe was recorded at oscillation amplitudes exceeding the sliding threshold at the dithering frequency of 70 Hz which is well below the resonance of the cantilever. As a result, the shear signal detected by a lock-in amplifier, was directly proportional to the maximal shear force in the junction – this uncalibrated signal is labelled 'shear response' in Fig. 2. The time constant for the thermal response detection was below 1 ms, and did not show any change due to application of dithering, nor any notable variation at this frequency, as expected due to the movement of the tip parallel to the sample surface. The recording of both shear and thermal responses was performed at a much slower rate of approximately 10 sec cycle, as the probe was gradually at 10 nm s$^{-1}$, brought into contact with the sample and then gradually retracted in the standard force spectroscopy way (26). These experiments were performed in ambient conditions on a 100 nm Au-coated polished quartz substrate (*SI, Fig. S1*).

Fig. 2 clearly shows that both thermal (curve a) and shear (curve c) responses have a strong dependence on the normal force. The normal force almost perfectly linearly increases as the sample is approached to and then pushed against the tip mounted on an AFM cantilever (curve b). In each case shown in Fig. 2 the zero *z* position corresponds to the onset (red line) of the solid-solid contact, whereas the breaking of the contact (blue line) appearing at a negative position of the sample surface, due to need to apply a negative (pull-off) force to overcome attractive forces acting between the tip and the surface.

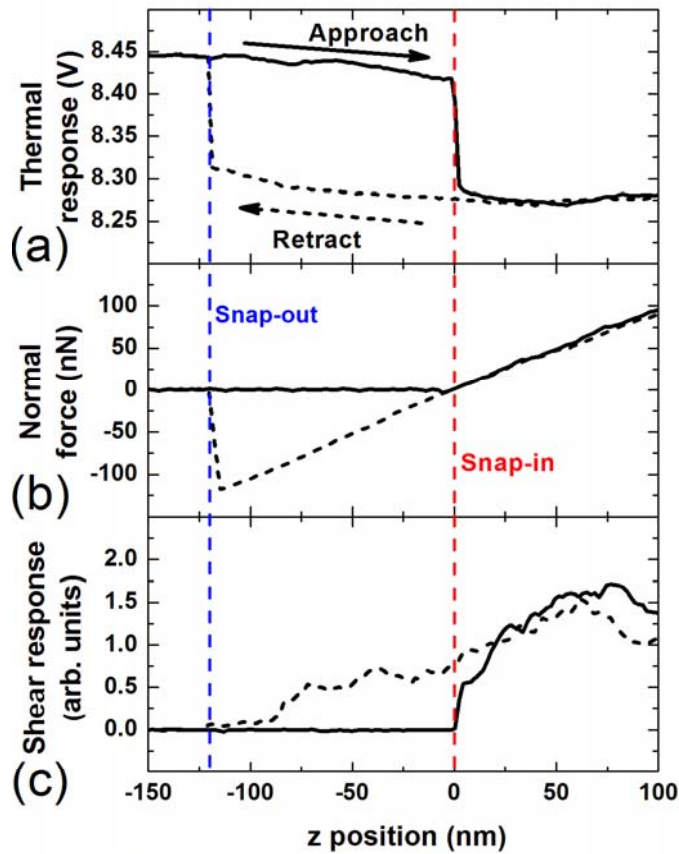

Fig. 2. Typical experimental single parameter investigation of probe-sample contact showing a, thermal response, a linear function of the tip-sample thermal resistance; b, normal force and c, shear response as a function of relative sample *z* - position of SThM probe on a polished quartz substrate. Solid lines correspond to the cantilever approach and dashed lines to the cantilever retraction. d, Correlation analysis (from 5 to 50 nm) between experimental thermal and shear responses, where the grey shaded region corresponds to simulated random responses (*SI text, Correlation analysis of thermal and shear responses*).

These studies indicated the presence of fine features observable in both the thermal contact fluctuations and in the shear force, with such features practically absent in the normal force response. Remarkably, the behaviour of the thermal resistance and shear response was clearly opposite, supported by the correlation analysis of these curves. The correlation coefficient between them was in the range of − 1 to − 0.8 (Fig. 2d), well above the level of ± 0.15 for the typical shear force and probe thermal response correlated with simulated noise (*SI, Correlation analysis of thermal and shear responses*). Reproducibility of the anti-correlation relationship between shear and thermal response has been confirmed across a wide range of samples: gold (metal with thermal conductivity ~310 W m$^{-1}$ K$^{-1}$) silicon (semiconductor with ~149 W m$^{-1}$ K$^{-1}$) and quartz (insulator with thermal ~1-10 W m$^{-1}$ K$^{-1}$). This correlation was shown to be valid for both types of most widely used SThM probes - Si based (DS) and Si$_3$N$_4$ (SP) probes (*Methods*).

**Determination of heat transfer pathways:** Quantitative heat transport analysis in SThM in ambient conditions is quite challenging as the probe temperature depends on multiple heat transfer pathways each with its own associated thermal resistance: a) solid-solid contact of the probe apex and the sample (27), b) through-the-air heat conduction (28), c) heat transfer via liquid bridge of condensed water in the contact region (29), and d) radiative far-field and near-field heat transfer, the latter has been shown to be generally insignificant in most SThM measurement conditions (8). To investigate the relative importance of the heat transfer pathways a-c) and therefore single out the nanoscale heat transfer phenomena at the tip-sample nanoscale junction, we conducted detailed comparative SThM studies in air and high vacuum (at $1 \times 10^{-7}$ Torr pressure) environments. Here a DS microfabricated probe was brought into contact with a polished Si surface while both thermal response and shear response was monitored during approach-retract cycles. In order to quantify the link between thermal and shear force phenomena, we now will compare the shear response with the thermal conductance of the probe-surface junction, $G_j$, inversely proportional to the junction thermal resistance such that $G_j=1/R_j$.

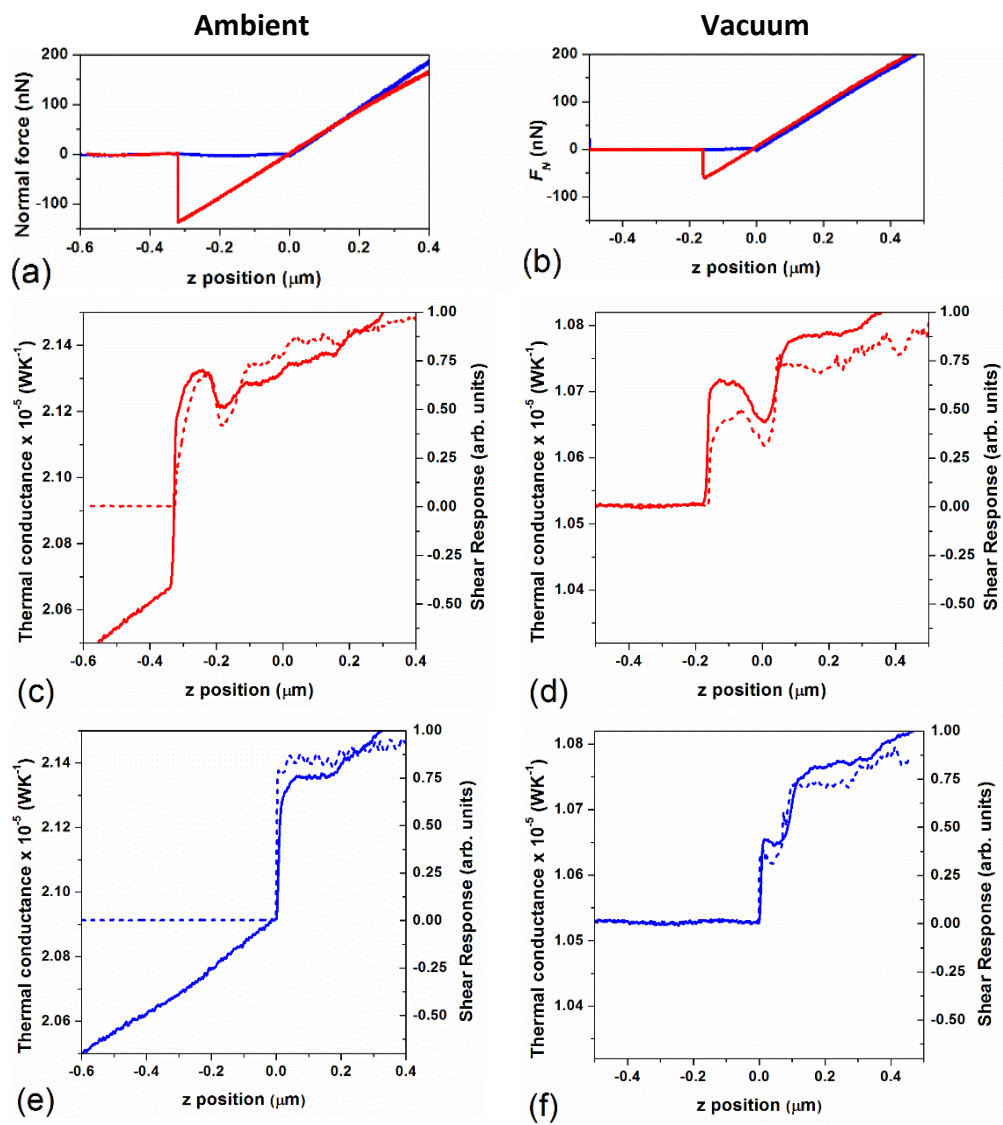

**Fig. 3.** Approach-retract cycles in (a, c, e) ambient and (b, d, f) vacuum environments using DS probe. c-f, Simultaneously measured total thermal conductance of the probe $G_p$ (solid lines) and maximal lateral force (dashed lines) values obtained during tip retracting from (c, d) and approaching to (d, f) to the surface.

The gradual change of the probe temperature as the tip approached the surface was clearly observed in air. Such change is associated with gas conduction of heat from the cantilever to the surface (Fig. 3e) and is fully eliminated in high vacuum (HV) environment (Fig. 3f). Furthermore, as expected, in air there was significantly greater adhesion due to the presence of the tip-sample water bridge; pull-off forces of ~137 nN and ~62 nN in air and in vacuum, respectively, were measured. Such water bridge should be completely eliminated in HV conditions of $10^{-7}$ torr.

The thermal contact in the nanoscale junction measured *via* SThM in vacuum, can be evaluated *via* the total probe thermal resistance, $R_{Total}$, arising from the probe thermal resistance ($R_p$, where the heat generated at the probe apex dissipates through the base of the probe) and a tip-sample junction thermal resistance, $R_j$, acting in parallel (30, 31), hence

$$R_{Total} = \frac{R_p \times R_j}{R_p + R_j}. \quad (1)$$

$R_{Total}$ is equivalent to the SThM thermal response in vacuum during solid-solid contact and $R_p$ corresponds to the SThM thermal response in vacuum immediately prior to snap-in to the sample. Measuring both $R_{Total}$ and $R_p$, allows $R_j$ and, hence, $G_j=1/R_j$, to be directly calculated. This allowed us to estimate the vacuum snap-in junction conductance at 65% of the in air values and therefore to confirm that the water meniscus, that is absent in HV conditions, cannot be regarded as the dominant heat conductance channel in the nanoscale-sized Si SThM probe (DS) contacts studied here. Furthermore, this suggests that the water meniscus for the sharp SThM probes is a less influential thermal pathway than was previously suggested (8, 31). In Fig. 3, we also observe that while one can broadly conclude that increase of normal forces produces a general rise of the probe thermal conductance, the shear response shows a clear and detailed correlation with the thermal conductance $G_p$.

## Discussion

For the simplest case of a single asperity solid-solid contact, one can find the relation between the force acting on the tip, observed as cantilever deflection (26), and the true probe-sample contact area *a* using the Johnson-Kendal-Roberts (JKR) model, this assumes that surface forces are short range in comparison to the elastic deformations they cause (32). This model is applicable here due to the relatively large adhesion forces and large tip radii. According to the model, the variation of contact area with normal force equivalent load (*L*) is given by $a/a_0 = ((1 + \sqrt{(1 - L/L_c)})/2)^{\frac{2}{3}}$ where $L_c$ is the negative critical load. *L* is derived from z-displacement using Hooke's law in the standard way (26), $L_c$ can be experimentally determined as the pull off force of the probe from the sample and $a_o$ is the contact radius at zero load such that $a_0 = ((6\pi\gamma r^2)/K)^{\frac{1}{3}}$ where *r* is the tip curvature radius, *γ* is interfacial energy per unit area (work of adhesion) and *K* is the combined elastic modulus of tip and sample such that $K = 4/3 \, ((1 - v_1^2)/E_1 + (1 - v_2^2)/E_2)$ where $E_1$ and $E_2$ are the Young's moduli of tip and sample, respectively, and $v_1$, $v_2$ are Poisson ratios of the tip and sample, respectively. Here $E_{SiO2}$ = 55 GPa and $v_{SiO2}$ = 0.27 (4). We can calculate *γ* directly from experimental measurement of $L_c$ such that $L_c = -\frac{3}{2}\pi\gamma r$ allowing the model to be pinned directly to the experimental results through the measured pull-off force on the probe. This approach reduces errors that may arise from, for example, surface contamination of the sample which could significantly change the value of *γ*. In this way, for our system comprising Si in vacuum, we calculate γ in the range 2.589 – 0.259 Jm$^{-2}$ for *r* = 5 – 50 nm, which compares well to 0.34 Jm$^{-2}$ in ambient conditions(4) previously reported for the tip with radius of curvature *r* = 10 nm. In summary, *r* and $L_c$ allow the calculation of γ and $a_0$ which in turn allows the calculation of *a* for different values of *L*. Here we derive *L* and $L_c$ from experiment and treat *r* as a variable.

Assuming the same single asperity geometrical symmetry between a tip and planar sample, it is generally assumed that, in vacuum, maximal shear response (33, 34) in the junction, $F_f$, is given by a continuum model such that $F_f = \tau \pi a^2$ where τ is the constant interfacial shear strength (21), it should be noted, however, that the validity of this model at the nanoscale is still a subject of some debate (35, 36). Furthermore, the thermal resistance of the probe-sample interface can be calculated using the method outlined by Prasher(37). Here we consider a general solution, defined by the Knudsen number *Kn* =Λ/*a*, where Λ is the mean free path of phonons in the material and *a* is the characteristic dimension, allowing to merge both the diffusive (*a* >> Λ) and ballistic (*a* ≈ Λ) case. The thermal resistance of the junction $R_j$ in vacuum can be written as $R_j = \frac{1}{2}ka\left[1 + \frac{8}{3\pi}Kn\right]$ where material thermal conductivity $k = \frac{1}{3}Cv_g l$, with *C* being the specific heat capacity per unit volume and $v_g$ is the phonon group velocity (here we use *C*=1.66 × 10$^6$ J/m$^3$ K and $v_g$= 6,400 m/s, respectively, for Si) (38). In the ballistic approximation, we assume that *Kn* →∞ and for ballistic heat flow regime junction resistance $R_{jb} = 4l/3k\pi a^2$, whereas for the diffusive regime at *Kn* →0, the junction thermal resistance becomes $R_{jd} = 1/(2ka)$. Hence both $R_{jb}$=1/$G_{jb}$ (37) and $F_f$ are related directly to the *actual* contact area (*A*=π*a*$^2$) between the probe and sample such that in ballistic regime,

$$R_{jb} = \frac{4}{Cv_g \pi a^2} \text{ and } F_f = \tau \pi a^2 \quad (2)$$

The detailed behaviour of these models as a function of r, and *L* is shown in *SI, Fig. S3*. This equation allows the comparison of two physical values that were measured simultaneously in the nanoscale junction *via* our SThM setup – namely, the heat conductance and the shear strength in the same junction. As all values except the tip radius of curvature, *r*, were experimentally measured, here we compare the experimental and modelled thermal conductance of the junction $G_j$ in vacuum (from data represented by Fig. 3 d,f) with the value of the maximal shear force in the junction ("shear response") as a function of *r*, in both ballistic (Fig. 4a) and diffusive (Fig. 4 b) approximations. The shear response has been normalised to the range 0-1 for the same maximum load (0 equivalent to absence of friction immediately prior to snap in to contact and, here, 1 equivalent to $F_f$ at *L* = 150 nN) to allow direct inter-comparison between series of measurements without the need to fully characterise the shear strength τ.

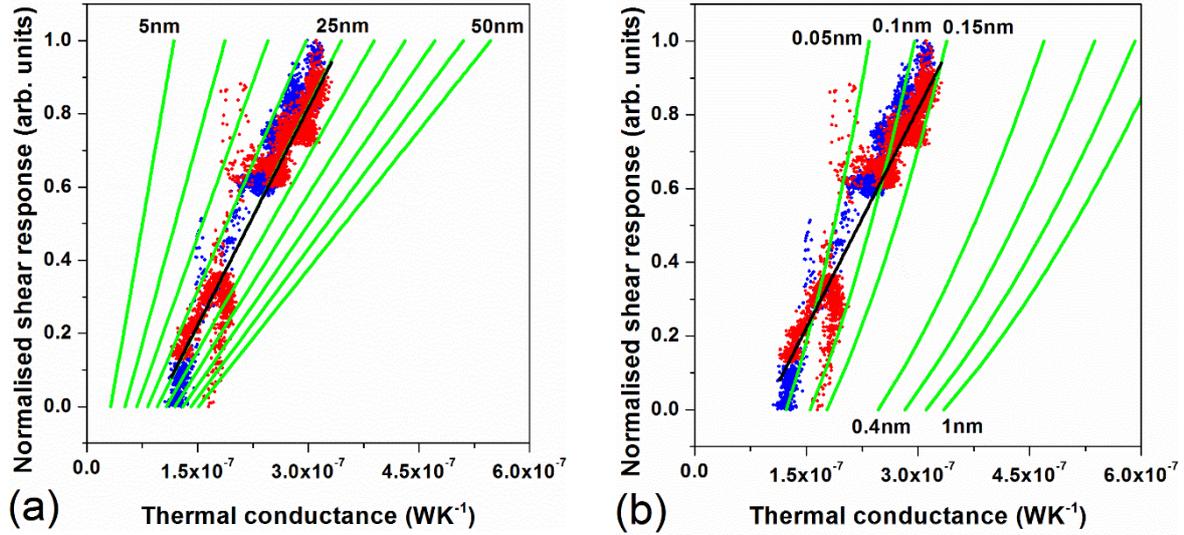

Fig. 4. Comparison of experimentally measured values of maximal shear force in the junction and junction heat conductance $G_j=1/R_j$ for approach (blue) and retract (red) data with the model (green) for of single asperity contacts. a, the ballistic ($R_{jb}$) and b, diffusive thermal ($R_{jd}$) approximation. A black line is a linear fit to the average of the experimental data, green lines show the modelling data for a series of single asperity contacts incremented in the range $r$ = 5-50 nm for $R_{jb}$ and at values $r$ = 0.05, 0.1, 0.15 and 0.4-1nm $R_{jd}$.

In the ballistic limit, the total contact area between probe and sample corresponds to the equivalent of a single asperity contact formed by a probe of $r$ = 25 nm which matches well the manufacturers stated value of $r$ < 30 nm, with excellent fit for the all measured range of data for shear strength and the thermal conductance. At the same time, assuming the diffusive heat transport approximation, the experimental data could only be approximated at single point for $r$ ~ 0.1 nm. The absence of the fit across the measured range of shear strength and heat conductance and physically unrealistic radius of 0.1 nm directly confirms the ballistic nature of heat transport in the nanoscale heat junction where individual asperities are typically well below the MFP of the heat carriers for almost any physically realistic contact. Critically, knowing the value of $r$ allows the calculation of contact area $A$ for each point in the experimental data. It should be noted that this relationship does not hold for probe-sample interactions calculated using the DMT approximation (*SI, Fig. S4 and SI text, Modelling of shear and thermal responses*) and that the experimental ratio in Fig. 4 (black line), whilst generally linear, has some scatter and branching suggesting that the probe contact radius may vary during the measurement cycle. Therefore, while the general relationships can be confirmed with high degree of confidence, these details may be further investigated using a more detailed contact model using, for example, molecular dynamic simulations (36).

Starting with the single asperity case, in an analogy to the Wiedemann–Franz law (16) describing the relationship between the thermal conductivity and electrical conductivity of a metal (39), here we can relate the thermal conductance $G_j$ with the normalised shear force $F_f/\tau$ in a nanoscale contact between same material *via* fundamental parameters of group velocity and heat capacity

$$G_j = \frac{Cv_g}{4} \cdot \left(\frac{F_f}{\tau}\right) \quad (3)$$

Crucially, as the nanoscale junction operates in the ballistic limit of thermal transport, the MFP of phonons need not be considered. In the case of different contacting materials (*i=1,2*), as the force and shear strength are common for each pair, and the resistance of the probe and the sample are connected in series (9, 12), one would expect these to be an effective value and a function of $1/(Cv_i)$. Finally, as it is now the total area that governs both heat transport and shear forces, this relation should be equally valid for the generic case of multi-asperity contact that both depends on the area rather than the lateral dimension of the contact.

In conclusion, we elucidate heat transport in nanoscale solid-solid contacts by simultaneous monitoring of shear forces and the heat transport in the scanning thermal microscopy based approach. We observed a clear correlation between the thermal conductance and the maximal shear forces in the junction as normal load was varied during establishing and breaking nanoscale contact. Our analysis suggests that the heat transport in these typical nanoscale solid-solid contacts is ballistic in nature *via* single or multi-asperities contacts. Moreover, we were able to propose a generalised relationship between a continuum model of shear response and thermal conductance for nanoscale contacts smaller than the MFP of the sample which can be easily realised in practically any nanoscale contact pairs having phonon dominated heat transport and may play a significant role in improving the quality and reliability of measurement of nanoscale thermophysical properties and development of nanoelectromechanical systems.

**Materials and methods**

**SThM measurements**: Ambient SThM measurements were performed using a standard contact AFM setup (Bruker Multi-Mode, Nanoscope III) with 'half-moon' SThM probe holder (Anasys Instruments) using a Pd integrated heater (SP) probe (Kelvin Nanotechnologies). $1\times10^{-7}$ mBar high vacuum (HV) measurements were performed in a dedicated low vibration chamber using a HV compatible AFM (NT-MDT multimode Solver SPM) and doped Si probe (DS) (AN-200, Anasys Instruments) see *SI text, Probe specifications*. As a heated SThM tip is brought in contact with the surface some heat flows into the sample cooling the tip and hence changing the electrical resistance of the sensor. The probe's electrical resistance was calibrated as a function of applied voltage and temperature using the method described elsewhere,(12) a linear dependence of the probe resistance on its temperature was always observed.(25) In both ambient and vacuum cases, during measurement, the thermal probe represents part of a balanced Wheatstone electric bridge with a 4 $V_{AC}$ signal at 91 kHz frequency, provided by a precision function generator (Model 3390, Keithley instruments), with unbalance AC signal measured *via* lock-in amplifier (SRS-830, Stanford Research Systems). A constant power was applied to the sensor generating Joule self-heating, with the tip temperature measured simultaneously hence measuring probe thermal resistance (30).

**Shear force measurements:** A shear piezo actuator (Physik Instrumente, PI-121.03) was used to laterally oscillate (at 70 Hz frequency and a drive voltage of 0-5 V) the sample perpendicular to the longitudinal axis of the cantilever during probe approach-retract. In-phase (x) and out-phase (y) component of the friction response were recorded by a lock-in amplifier (SRS-830, Stanford Research Systems). The lateral dithering frequency is selected to be several orders of magnitude below flexural and torsional resonances of the cantilever, and the frequency response of the shear actuator. The dithering amplitude was above the sliding friction threshold, resulting in the torsional response of the cantilever directly proportional to the shear strength of the sliding contact. Given the cantilever torsional stiffness on the order of 100 Nm$^{-1}$, the shear force on the order of 1-10 nN, and the typical tip length for the probe used on the order of 10 μm[16], the torsional angle variation being less than 1-2 ×10$^{-5}$ rad, suggesting practically constant angular orientation of the tip apex and change in the tip-surface distance in the area of the contact of 10 nm on the order of 10$^{-13}$ m, with no effect of dithering on the heat transport being observed.

**Acknowledgements.**

We thank Dr. Peter Tovee for help with the SThM instrumentation and Professor Vladimir Fal'ko for useful discussions. Authors also grateful to funding by the EU (projects FUNPROB, QUANTIHEAT) and EPSRC grants EP/K023373/1, EP/G015570/1.

# SUPPORTING INFORMATION

## Shear Forces and Heat Conductance in Nanoscale Junctions.

Benjamin. J. Robinson,[1,2] Manuel. E. Pumarol,[1] and Oleg. V. Kolosov*[1,2]

[1] Physics Department, Lancaster University, Lancaster, LA1 4YB, UK; email: o.kolosov@lancaster.ac.uk

[2] Materials Science Institute, Lancaster University, Lancaster, LA1 4YW, UK

**SI Text**

**Probe specifications.**

SP probe (Kelvin Nanotechnology), is made of low thermal conductivity silicon oxide or silicon nitride ($k_{SP}$ ~ 1-4 Wm$^{-1}$K$^{-1}$) and has a Pd resistive temperature sensor that is positioned close to the apex of the tip. Manufacturers stated tip radius is <100 nm but is typically found to be ca. 30-50 nm. DS probes (Anasys Instruments), are made of single crystal Si of high thermal conductivity ($k_{DS}$ ~ 130 Wm$^{-1}$K$^{-1}$) and a small radius of curvature of the tip of less than 30 nm. The highly doped cantilever leg provides a current path to a moderately doped resistive temperature sensing part of this probe, with the latter separated from the tip apex by the conical probe tip.

**Analysis of probe approach in ambient conditions**

During approach of the heated SThM probe, in ambient conditions, a sharp snap-in to contact occurs when the gradient of attractive forces, originating in van der Waals interaction and water capillary bridging, (1) between sample and probe is greater than the cantilever spring constant, $k$ (at $k_{SP}$ ~0.3 Nm$^{-1}$). Simultaneously, a sharp decrease in probe thermal resistance ("thermal response") at the surface (z position of 0 nm) arises from conductance between the heated probe and sample due to the onset of solid-solid contact. The gradient of the thermal response prior to snap-in (negative z position) is due to conduction through the decreasing probe-sample air gap (1, 2). For Au surface, immediately after the contact, the probe temperature decreases indicating decrease of the thermal resistance of the probe, due to increase of the contact area following the increase of the normal force. Simultaneously, the maximum friction force achievable in the junction ("shear response") was zero prior to solid-solid contact, but increased after the initial contact with the increase of the load (3). It should be noted that for SiO$_2$ sample, both the thermal and shear response post-contact are far from being monotonous (arrows in Fig.S1a) that may be linked with the particular shape of the probe apex in contact with the sample, so that a changing zone of the probe apex is contacting the surface as the probe approaches. We noted no change in response over multiple (~100) cycles indicating negligible wear of the probe or sample.

**Correlation analysis of thermal and shear responses**

Correlation analysis, using the Pearson product-moment correlation coefficient, was performed on the multiple sets of experimental data of the thermal resistance and shear response for both Quartz and Au, with typical responses shown in Fig. S1. In each case a correlation coefficient between -1 and +1 is generated where -1 indicates a perfect negative correlation, +1 indicates a perfect positive correlation and coefficients around 0 indicates no correlation.

Here we compared experimental thermal data with randomly generated shear response data (Fig. S2a,b) of the same overall mean value and maximum/minimum values and the same number of measurement points. Similarly, we compared experimental shear data with randomly generated thermal data (Fig S2c,d). In both cases coefficients were typically in the range of $\pm 0.15$. Correlation coefficients between experimental shear and experimental thermal responses were in the range $-1$ to $-0.8$ (Fig. S2e,f) significantly statistically different from the comparison of correlation between one set of experimental data and the random data.

**Modelling of shear and thermal responses**

All modelling reported is based upon the Johnson–Kendall–Roberts (JKR) model of probe-sample interaction which assumes that surface forces are short range in comparison to the elastic deformations they cause (4) and is applicable here due to the relatively large adhesion forces and large tip radii. Fig S3 shows modelling of probe-sample responses and shear and thermal dependences using the JKR equations in the main manuscript.

If the opposite limit of the tip-surface properties (stiff probe with small radius of curvature), the probe-sample interaction can be described by the Derjaguin–Mueller–Toporov (DMT) model such that the variation of $a$ with normal force equivalent load ($L$) is given by $a/a_0 = (1 - L/L_c)^{\frac{1}{3}}$ where $L_c$ is the negative critical load is given by $L_c = -2\pi\gamma r$ and $a_o$ is the contact radius at zero load such that $a_0 = ((2\pi\gamma r^2)/K)^{\frac{1}{3}}$ where $r$ is the tip curvature radius, $\gamma$ is interfacial energy per unit area (work of adhesion) and $K$ is the combined elastic modulus of tip and sample. The equivalent DMT probe-sample responses and shear and thermal dependences are shown in Fig. S4.

Additionally, we compare the experimental and theoretical thermal and shear response derived using the DMT model of probe-sample interaction (Fig. S5), indicating that DMT model is not describing well the force interactions and heat transport in the nanoscale junction.

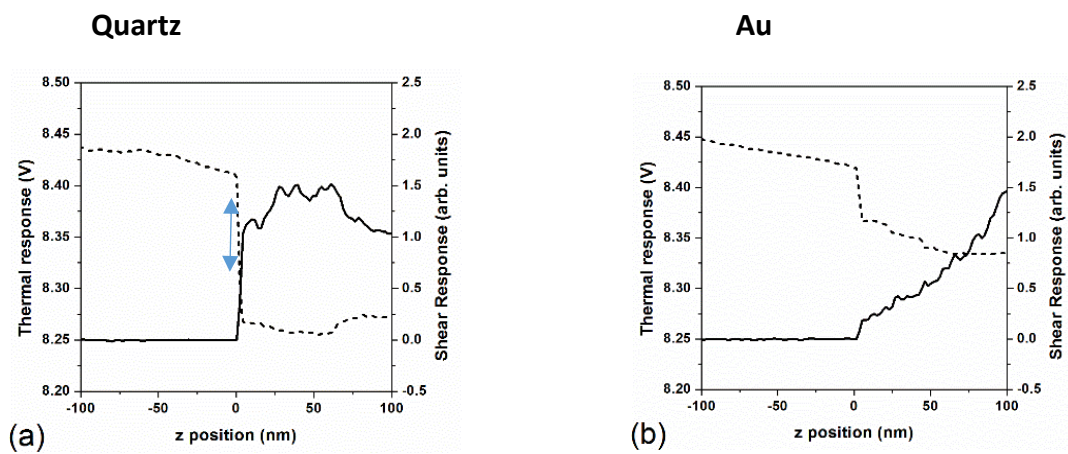

**Fig. S1.** Comparison of Au and Quartz thermal and shear responses showing two representative thermal (dashed lines) and shear (solid lines) responses of a nanoscale SP SThM probe approaching a, quartz and b, Au surfaces (z=0 corresponds to the sample surface).

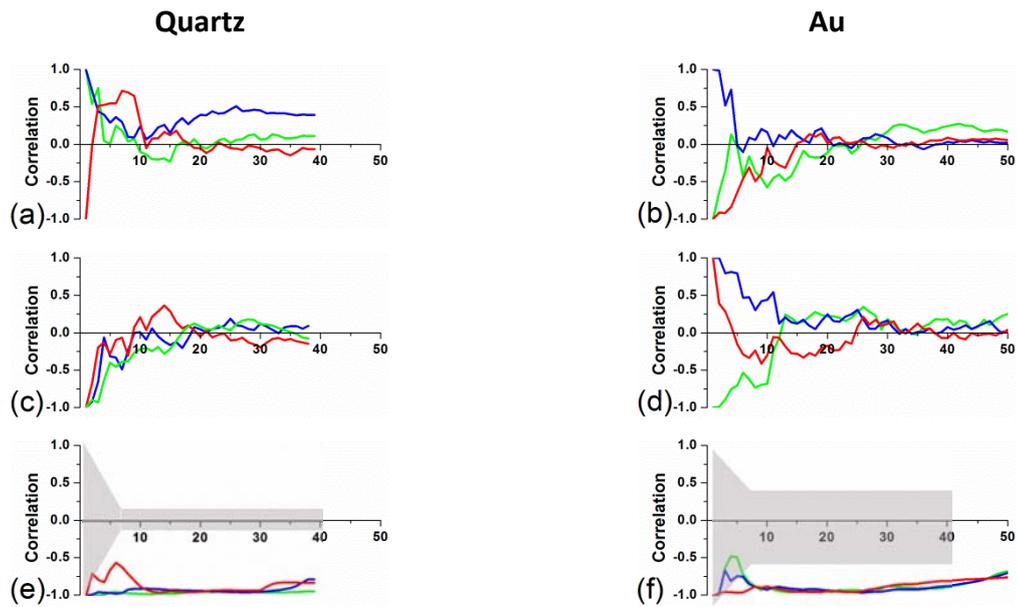

Fig. S2. Correlation analysis of simulated and experimental shear response and thermal resistances where a, b experimental thermal resistance data vs random distribution shear response data for quartz and Au respectively. c, d experimental shear response data vs random distribution thermal response data for quartz and Au respectively. e, f correlation of experimental data for both shear and thermal responses demonstrating a correlation coefficient in the range -0.8 to -1.0. In all cases correlation is calculated on the solid-solid contact regime corresponding to z position of 5 to 50 nm.

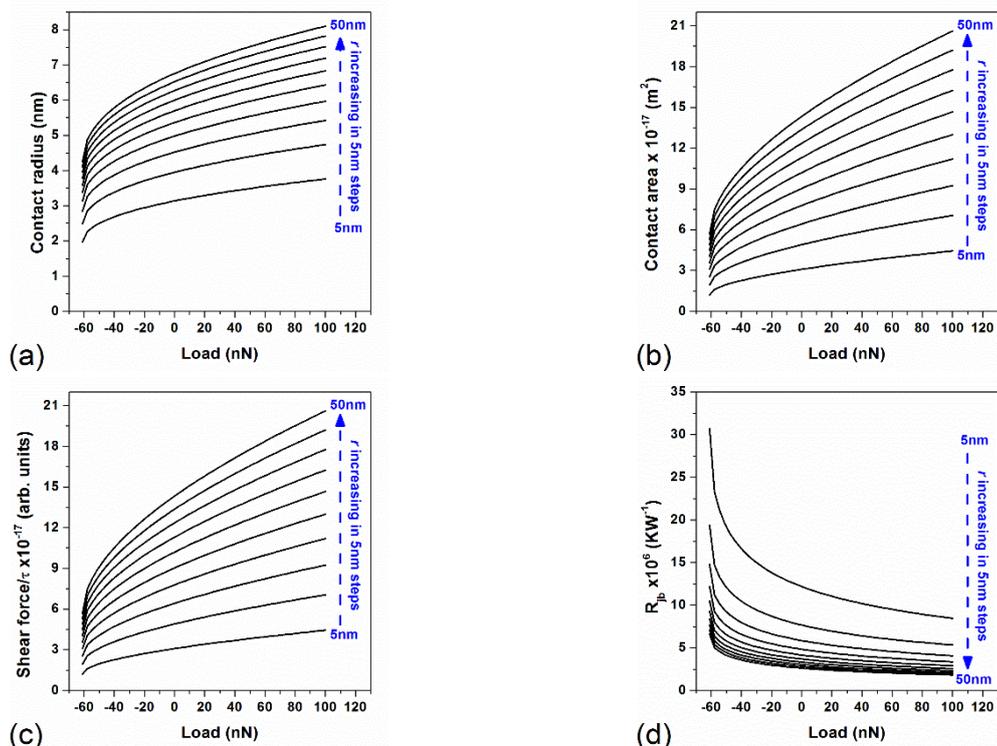

Fig. S3. JKR based modelling of actual contact geometry, shear and thermal responses in vacuum showing calculated a, actual contact radius b, actual contact area c, shear force/$\tau$ and d, contact thermal resistance in the assumption of ballistic heat transport ($R_{jb}$), corresponding to the experimental system in vacuum as function of load $L$ for a probe radius of curvature $r$.

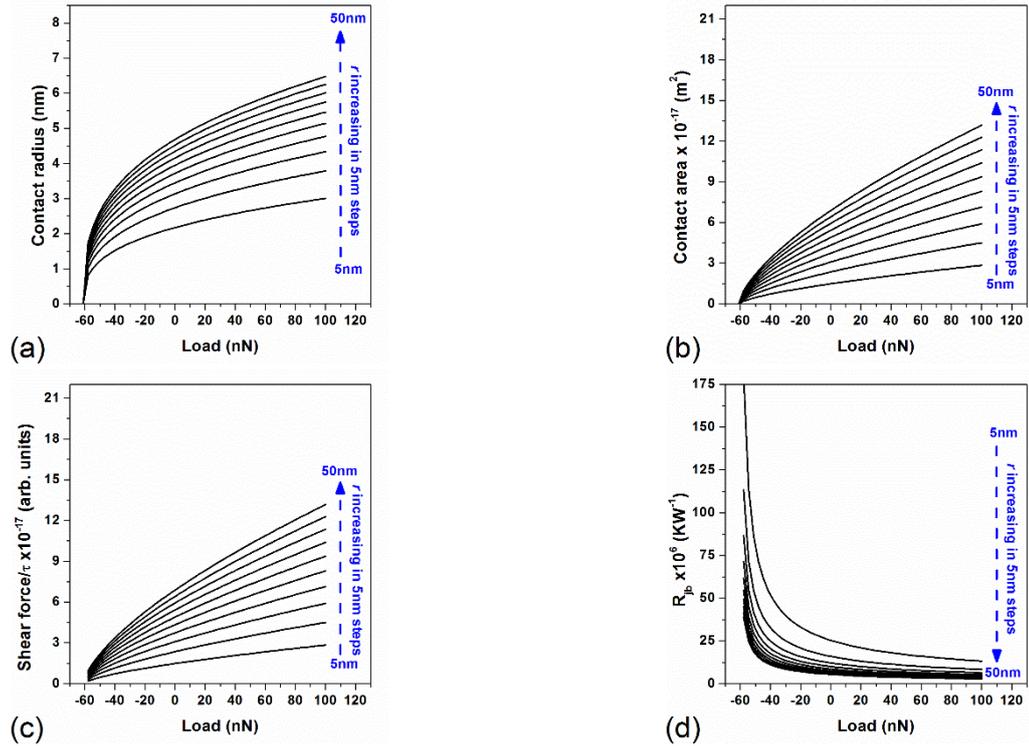

**Fig. S4.** DMT based modelling of actual contact geometry, shear and thermal responses in vacuum showing calculated a, actual contact radius b, actual contact area c, shear force/τ and d, contact thermal resistance in the assumption of ballistic heat transport ($R_{jb}$), corresponding to the experimental system in vacuum as function of load L for a probe radius of curvature r.

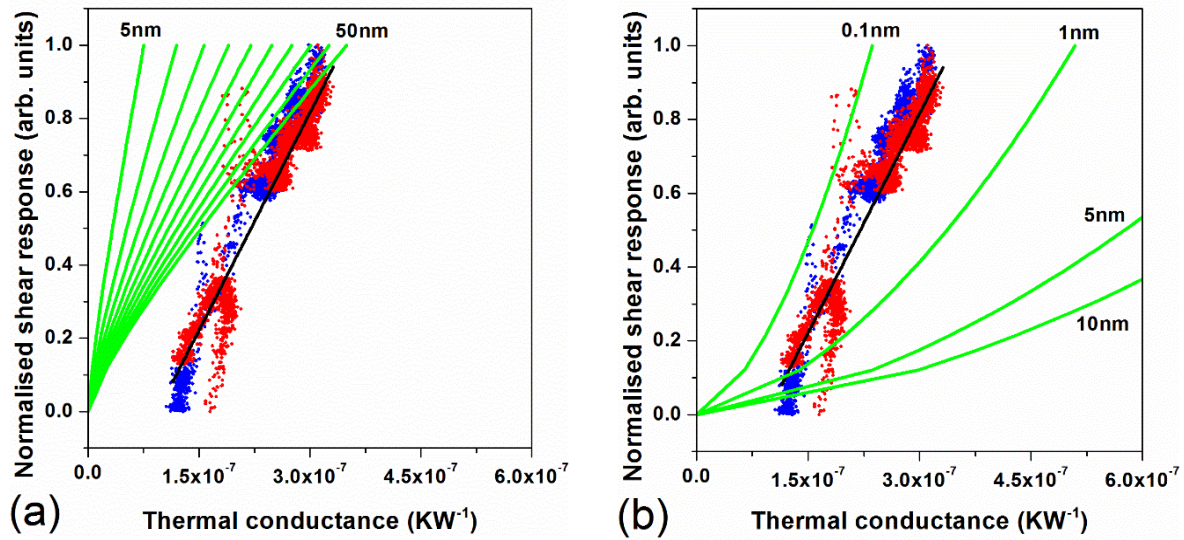

**Fig. S5.** Comparison of experimental approach (blue) and retract (red) data and modelling (green) for thermal responses of single asperity contacts in a, the ballistic and b, diffusive thermal limits. A black line is a linear fit to the average of the experimental data, green lines show the modelling data for a series of single asperity contacts incremented in the range r = 5-50 nm for the ballistic model and at values r = 0.1, 1, 5 and 10 nm the diffusive model.